%% file: qcd20.tex
\documentclass[aps,prd,preprint,groupedaddress,superscriptaddress,nofootinbib,11pt
]{revtex4-1}

\usepackage{graphicx}
\usepackage{dcolumn}
\usepackage{bm}

\usepackage{amsfonts}
\usepackage{amsmath}
\usepackage{graphicx}
\usepackage{hyperref}
\usepackage{anyfontsize}
\usepackage{ulem}
\usepackage[usenames]{color}
\hypersetup{colorlinks, linkcolor = [rgb]{0,0.0,0.5}, citecolor = [rgb]{0,0.0,0.5}, urlcolor = [rgb]{0,0.0,0.5}}

\input{lan-def}

\begin{document}

\preprint{SLAC-PUB-17374}

\title{Exotic  States in a Holographic Theory}
\thanks{Talk given at 23rd International Conference in Quantum Chromodynamics (QCD20),  27-30 October 2020, Montpellier, France.}%

\author{H. G. Dosch}
\thanks{Speaker, corresponding author}
 \email{h.g.dosch@gmail.com}
 \affiliation{Institut f\"ur Theoretische Physik, der Universit\"at Heidelberg, Philosophenweg 16, 69120 Heidelberg, Germany}

\author{S. J. Brodsky}%
\email{sjbth@slac.stanford.edu}
\affiliation{SLAC National Accelerator Laboratory, Stanford University, Stanford, CA 94309, USA
}%


\author{G. F. de T\'eramond}
 \email{gdt@asterix.crnet.cr}
\affiliation{ Laboratorio de F\'isica T\'eorica y Computacional,
 Universidad de Costa Rica, 11501 San José, Costa Rica
}%

\author{M. Nielsen}
\email{mnielsen@if.usp.br} 
\affiliation{Instituto de F\'isica, Universidade de São Paulo, 05508-090 São Paulo, São Paulo, Brazil}%

\author{Liping Zou}
\email{zoulp@impcas.ac.cn}
\affiliation{Institute of Modern Physics, Chinese Academy of Sciences, Lanzhou 730000, China}


\begin{abstract}

Supersymmetric Light Front Holographic QCD is a holographic theory, which  not only describes the spectroscopy of mesons and baryons,  but also predicts the existence and spectroscopy of tetraquarks.  A discussion of the limitations of the theory is also presented.

\end{abstract}

\maketitle

\section{Introduction}

Light front holographic QCD~\cite{Brodsky:2003px,Brodsky:2006uqa,deTeramond:2008ht,Brodsky:2014yha} (LFHQCD) is a bottom-up holographic theory.  Supersymmetric LFHQCD is a semiclassical model for hadron physics based on the equivalence~\cite{Maldacena:1997re} of a classical 5-dimensional theory (AdS$_5$) and a superconformal 4-dimensional quantum field theory (AdS/CFT) and on quantization in the light front (LF)  frame. The breaking of the conformal symmetry, necessary for the generation of a discrete mass spectrum, is achieved by assuming that the resulting LF Hamiltonian is a linear combination of generators of the superconformal  graded algebra~\cite{Witten:1981nf,Fubini:1984hf}. Following this procedure the form of the interaction is completely fixed and, in the limit of massless quarks, it is determined by one  dimensionful constant $\la$,  representing the QCD confining interaction. The spin dependence is taken from the tensor structure of the fields in the in the 5-dimensional classical theory~\cite{deTeramond:2013it}, and the quark mass dependence from  mapping to the LF Hamiltonian. In this way the number of parameters is minimal, as in lattice gauge theory.  Supersymmetric LFHQCD theory is not a supersymmetric field theory with squarks and gluinos, but a semi-classical theory with supermultiplets of hadrons.  Supersymmetric LFHQCD describes not only hadronic spectroscopy, but it gives also insight into the dynamical and structural properties of the strong interactions~\footnote{For a recent recapitulation see~\cite{Brodsky:2020ajy}.}.

 \section{Supersymmetric Light Front Holographic QCD} 
 
 As mentioned above supersymmetric LFHQCD is based on broken superconformal quantum mechanics~\cite{Fubini:1984hf,deTeramond:2014asa, Dosch:2015nwa}.
The Hamiltonian (translation operator in time) $H$ of superconformal quantum mechanics  has the same form as that of the light-front Hamiltonian derived from AdS$_5$. It is the anti-commutator of a fermionic operator $Q$, $H=\{Q,Q^\dagger\}$. Another fermionic operator $S$ of the graded superconformal algebra is the ``square root of the special conformal transformation'' $K$:  $K= \{S,S^\dagger\}$. In a two-component Hilbert-space $\cL_2(\cR_1) \oplus \cL_2(\cR_1)$, where the upper component is defined as the bosonic component and the lower component as the fermionic one, the two fermionic operators can be represented as: $Q=\left(\begin{array}{cc} 0 &-\pa_x+\frac{f}{x}\\
0&0\end{array}\right), ~ S=\left(\begin{array}{cc} 0 & x\\ 0&0\end{array}\right)$. 
In supersymmetric LFHQCD a scale is introduced by changing the Hamiltonian $H$ from $H=\{Q,Q^\dagger\}$ to
 \beq 
 \lb{eG}G= \{R_\la,R_\la^\dagger\} \quad \mbox{with} \quad R= Q+ \la S. 
 \enq
 
The operators  $Q$ and $S$ have  different dimensions and, therefore, the linear combination of the two must contain a dimensionful constant. This constant $\la$ in  (\ref{eG}) with the dimension of a squared mass is the scale of supersymmetric LFHQCD. 

The resulting hadron mass spectra of the Hamiltonian $G= \{R_\la,R_\la^\dagger\}$ for mesons and baryons are:
\beqa \lb{e1}
M^2 _M&=& 4\, \la \, (n+L + S/2)   + \De(m_1^2,m_2^2),   \\
M_B^2&=& 4\, \la \, (n+L + S/2+1)   + \De(m_1^2,m_2^2,m_3^2), \nn
\enqa
with internal spin $S = 0, 1$.
The parameters from a fit of all light hadron spectra come out to be:
$ \sqrt{\la} = 0.523 \pm  0.024$  GeV and the effective masses for the up and down quarks are 
$m_q=  0.046$ GeV, $q = u,d$, and for the strange quark   $m_s=0.357$ GeV. 
Thus, supersymmetric light-front holography explains the remarkable degeneracy of the meson and baryon Regge trajectories.

The analytical expressions for the $\Delta$-terms in Eqs.~\req{e1}, which describes the influence of the quark masses,  is~\cite{Brodsky:2014yha,Dosch:2015nwa,Dosch:2015bca,Dosch:2016zdv,Brodsky:2016yod} 
$$ \De M_n^2(m_1, \cdots
m_n)=  {-2 \la^2}  \frac{\pa}{\pa \la}\log K(\la), $$
with
$$K(\la)=  \int_0^1 dx_1
\cdots dx_n \, \de(x_1 + \cdots x_n -1) \,  e^{- \frac{1}{\la} (\frac{m_1^2}{x_1} +\cdots \frac{m_n^2}{x_n})}. $$

The holographic equivalence is based on a large number of colours~\cite{Maldacena:1997re} and therefore the accuracy in this $1/N_c \to \infty$ approximation is limited approximately to  $1/N_C^2$, thus to $\pm 120$ MeV.
In Fig. \ref{fig1}  the value of the dynamical scale $\sqrt{\la}\;$ is shown, as determined from the different hadronic and mesonic channels; within the expected accuracy it clearly indicates the meson-baryon supersymmetry.

\begin{figure} 
\includegraphics[width=0.7 \textwidth]{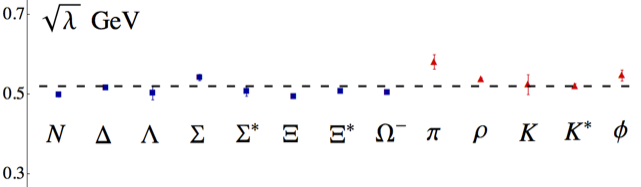}
\vspace{-0.1cm}
\caption{\lb{fig1} The dynamical scale $\sqrt{\la}  = 0.523 \pm 0.024 ~{\rm GeV}$ as determined from different hadronic and mesonic channels including orbital and radial excitations.}
\end{figure}

\begin{figure} 
\hspace{-1cm} \includegraphics[width=0.8 \textwidth]{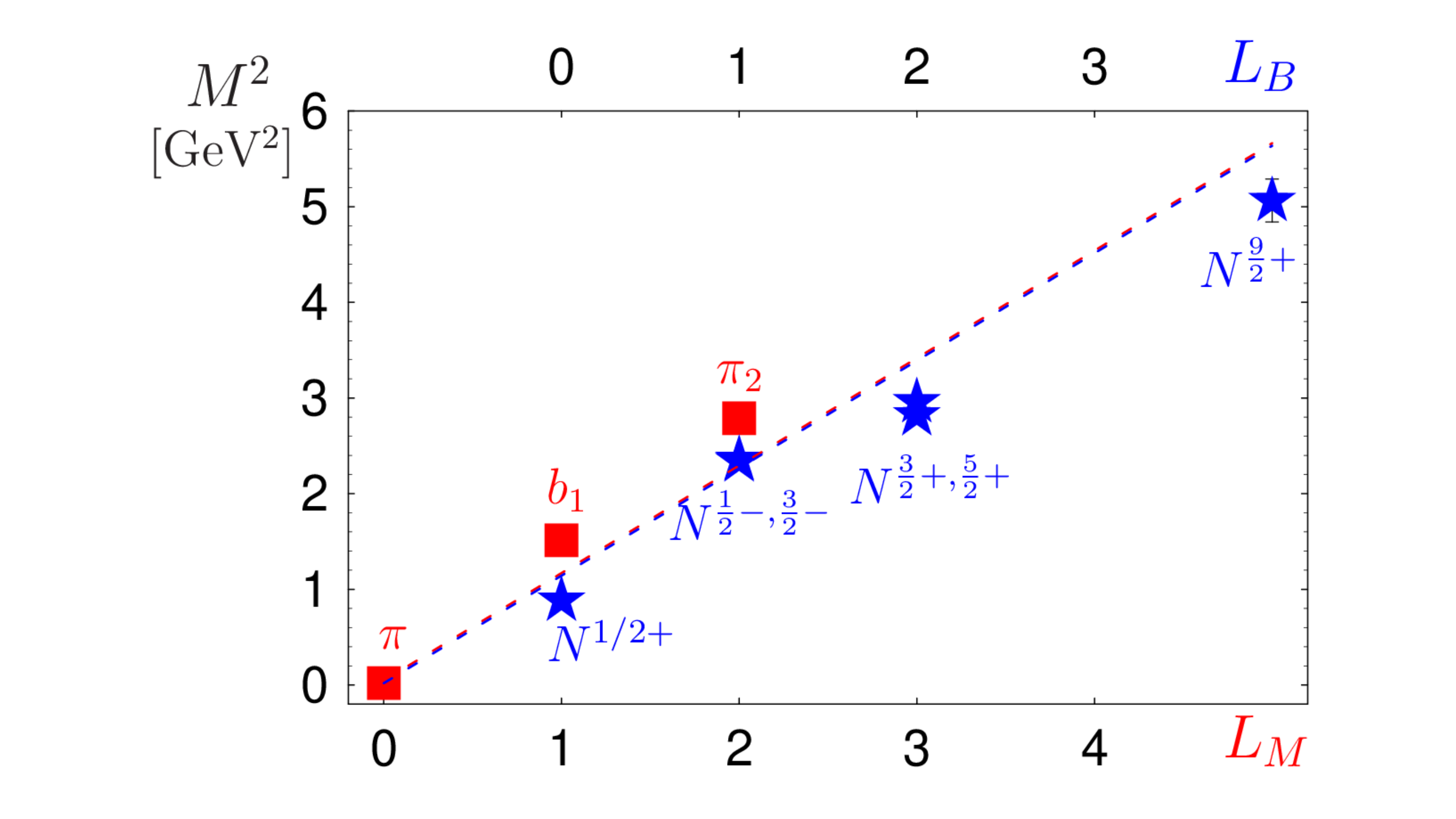}

\hspace{-1cm}  \includegraphics[width=0.8 \textwidth]{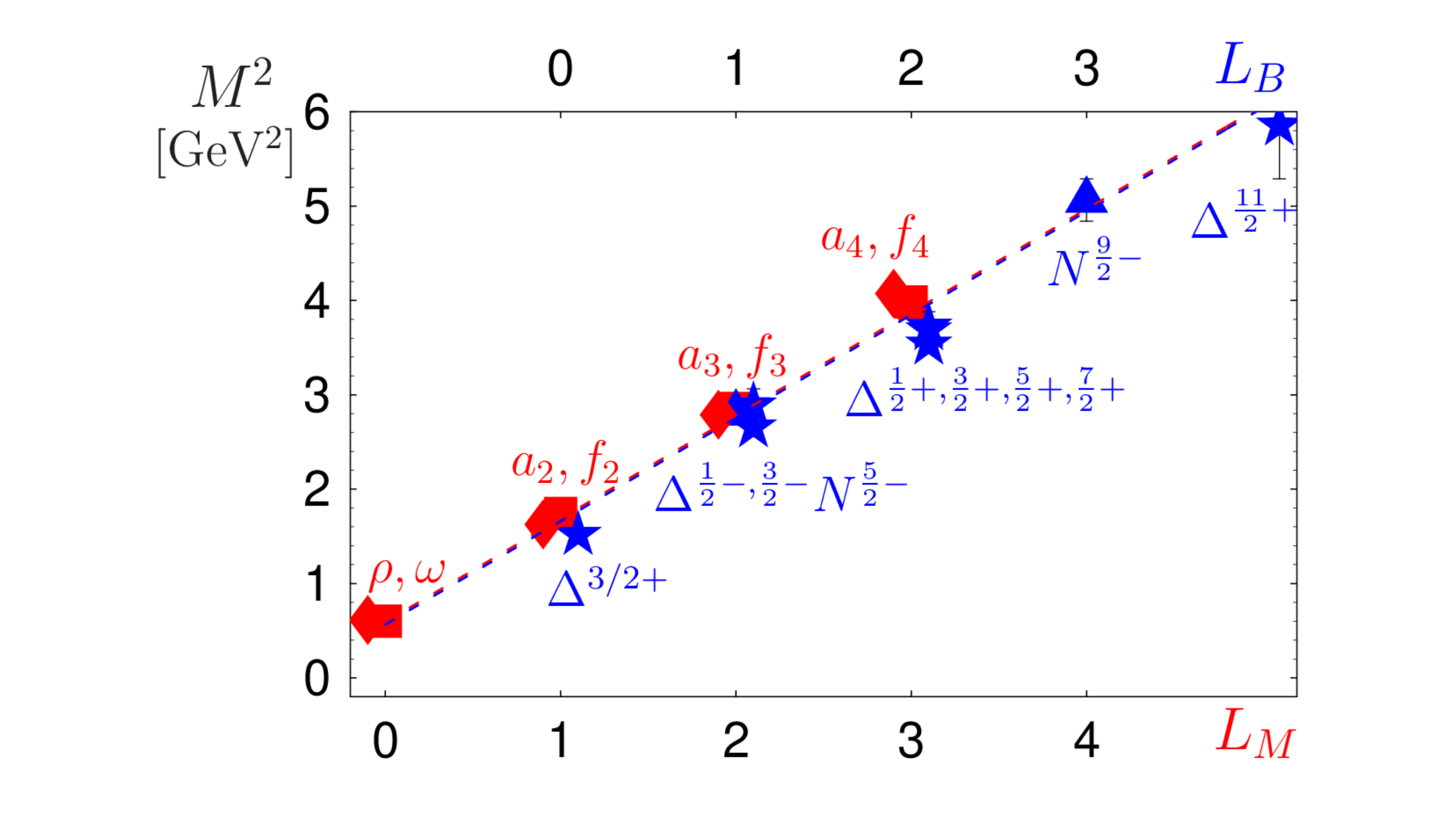}

\hspace{-0.65cm}\includegraphics[width=0.76\textwidth]{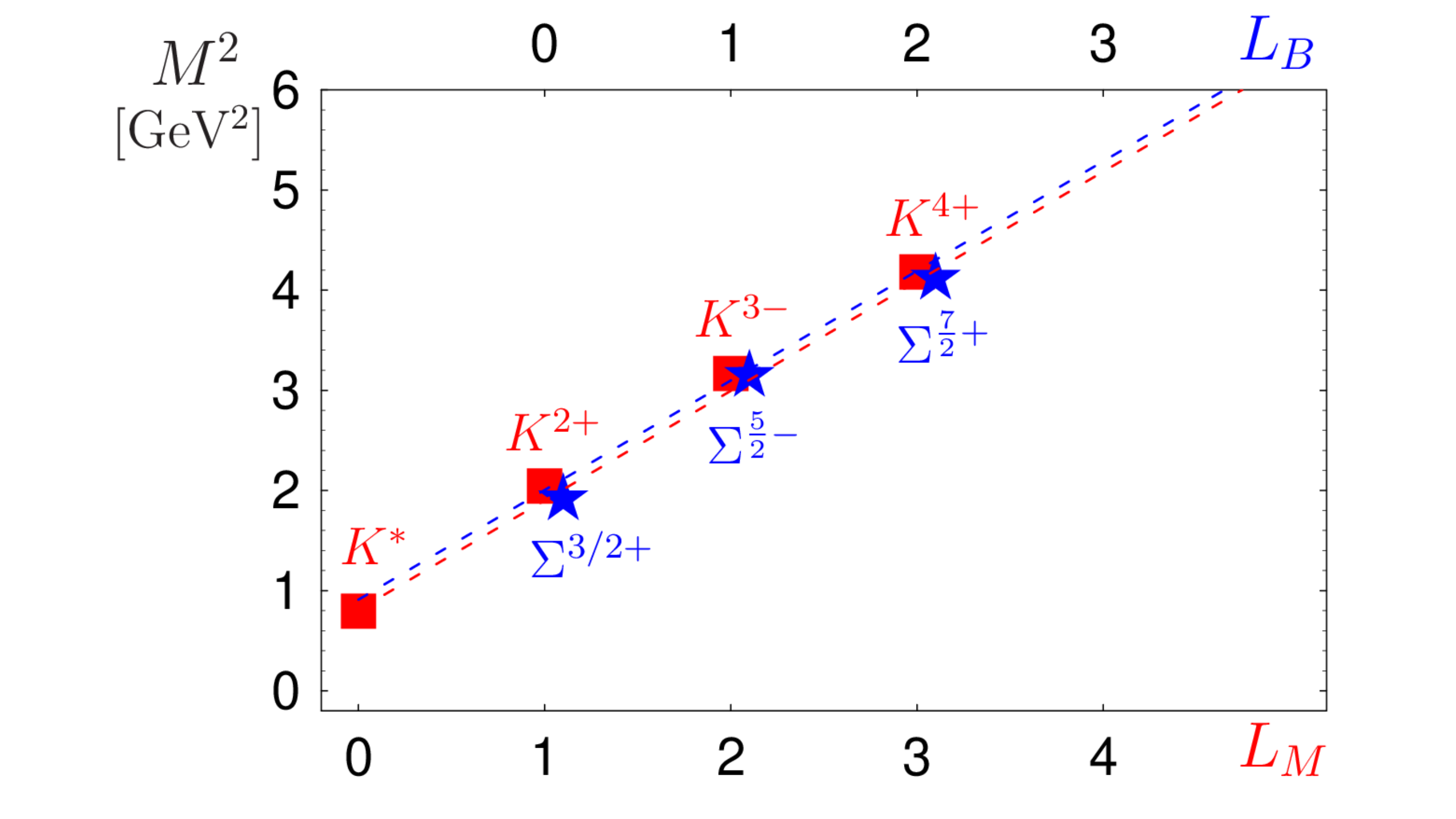}

\caption{\lb{fig2} Leading light mesonic and baryonic trajectories: The dashed curves are the predictions of supersymmetric LFHQCD, Eqs.~\ref{e1}.}
\end{figure}

In Fig. \ref{fig2} some leading meson and baryon trajectories are shown. The meson-baryon symmetry is evident. The lowest mesonic state, which in a chiral theory has zero mass, has no super-partner. In the supersymmetric theory it plays the role of a vacuum state~\cite{Witten:1981nf}.

\begin{figure} 
\includegraphics[width=0.8 \textwidth]{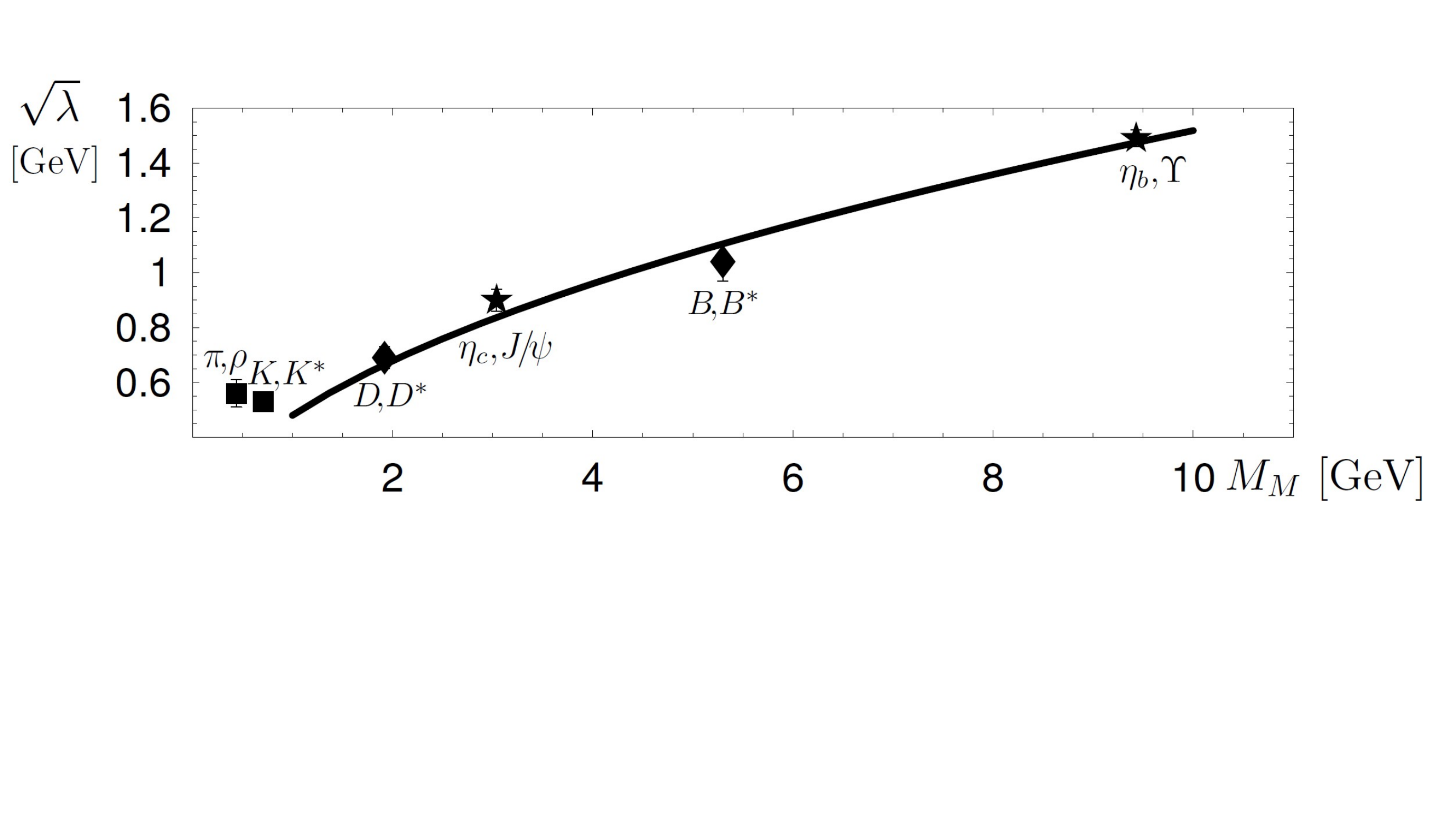}
\vspace{-3cm}
\caption{ \lb{fig3}The value of the dynamical scale $\sqrt{\la}$ as determined from the lowest lying mesonic sates  containing one or two heavy quarks. The solid line is the curve $\sqrt{\la_Q}= 0.49 \,\sqrt{M_M}$ [GeV].}
\end{figure}

For hadrons containing heavy quarks the  conformal symmetry  is strongly broken, but due to the constraints imposed by  supersymmetry \cite{Witten:1981nf} and heavy quark effective theory~\cite{Isgur:1991wq} the form of the interaction remains unchanged, ony the numerical value of the scale $\la$ 
changes from the values for light quarks from $\sqrt{\la} =0.523 $ GeV for heavy quarks to a value
\beq 
\lb{la-q} \sqrt{\la_Q}= C\,\sqrt{M_M},
\enq 
where  $C \approx 0.49 \, \sqrt{\rm GeV}$~\cite{Dosch:2016zdv} and  $M_M$ is the mass of the lightest meson containing the heavy quarks, see Fig. \ref{fig3}. There
are less data in this sector, but there is no experimental contradiction to the meson-baryon supersymmetry: Errors are somewhat larger, due to uncertainty of $\la_Q$, a rough (optimistic) estimate is $\De M \approx 150$ MeV.

\subsection{Completion of the Supermultiplet}

\begin{figure}
\includegraphics[width=1.0 \textwidth]{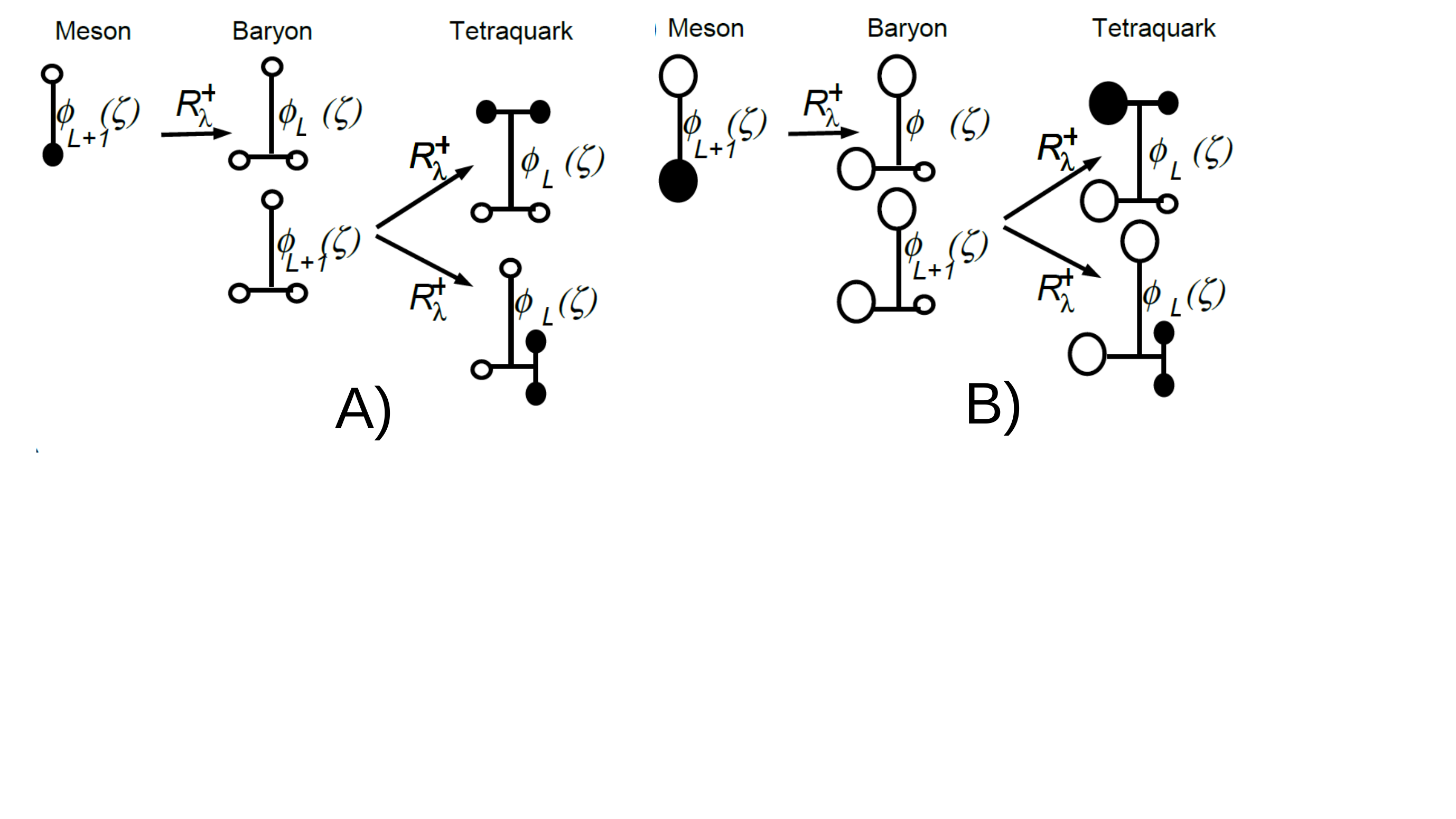}
\vspace{-5cm}
\caption{\lb{fig4}  The interpretation of superconformal quantum mechanics in terms of LFHQCD. The operator $R_\la^\dagger$ transforms a quark (open circle) or an antiquark (filled circle)  of the meson into pair of antiparticles of the same color, that is a quark into an anti-diquark cluster with color $\bf 3$ or an antiquark into a diquark cluster of color $\bar{\bf 3}$. In the same way a constituent of the baryon is transformed into a  two-particle cluster of the same color. Therefore the resulting hadron is a tetraquark, consisting of a diquark - anti-diquark cluster.}  
\end{figure}

Up to now we have supermultiplets consisting of one meson and two-baryon components, one  with positive and one with negative chirality: 
$\ph_{L_M}$ and $ \psi^+_{L_M-1} \,  ,\ps^-_{L_M}$.
Note that in LF quantization the LF angular momentum of the negative chirality state  $\psi^-$ is by one unit higher than that of the positive chirality state $\psi^+$. The parity is determined by the LF angular momentum of the leading twist: the positive chirality state~\cite{Brodsky:2014yha}.

In a complete supermultiplet the number of bosons components equals the number of fermion components, we thus miss an additional boson. This is indeed included in the formalism~\cite{Dosch:2015nwa,Brodsky:2016yod} and the mass of this additional state  is given by supersymmetric LFHQCD.
The fermionic operator $R_\la$, see \req{eG}, transforms a bosonic meson wave function with LF angular momentum $L_M$ into a a fermionic baryon wave function with  $L_B=L_M-1$ and positive chirality: $R_\la^\dagger \ph_{L_M}  = \psi^+_{L_M-1}$.  
  It also  transforms the negative chirality baryon  wave function $\psi^-_{L_M-1}$ of a baryon, which has leading twist LF angular momentum $L_M$, into a a bosonic wave function with  LF angular momentum $L_M-1$:
\beq
R_\la^\dagger \psi^-_{L_M} = \Ph_{L_M}.
\enq

This additional bosonic state $\Ph_L$ completes the  supersymmetric quadruplet of two mesons and two fermions  which can be arranged into a matrix 
 \beq  \lb{SM}
 {\Ps}= \left( \begin{array}{cc} \ph_L& \ps_L^-\\ \ps_{L-1}^+ & \Ph_{L-1} \end{array} \right)  \quad \mbox{ with }\quad {G} \, {\Psi } = M^2 \, {\Ps}.
 \enq
This additional bosonic state can, in terms of LFHQCD, be interpreted as a tetraquark. The argument  is given in Fig.~\ref{fig4} caption: It consists of a diquark with color $\bf 3$ and an anti-diquark with color $\bar{\bf3}$. The spin-statistics theorem limits the possible states in LFHQCD, a diquark composed of identical quarks in the lowest lying state must be symmetric under the combined spin and isospin transformation, since it is antisymmetric in color, namely in a $\bf 3$ or $\bar{\bf3}$ state. 

As mass of the tetraquark one obtains 
\beq
 \lb{masst}
M_T^2= 4\, \la \, (n+L + S/2+1)   + \De(m_1^2,m_2^2,m_3^2,m_4^2).
\enq
In Table \ref{tab1} the lowest lying light tetraquarks are displayed, for hadrons with total isospin 0 additional possible non-perturbative effects have to be taken into account, see \cite{Zou:2019tpo}. In accordance with the spectra of mesons and baryons we expect an accuracy of $\approx 120$ MeV. Possible candidates for a quadruplet are $a_2(1320)\; \De(1232)\;a_1(1260)$. 

\begin{table}
\caption{\lb{tab1} Predicted masses of light  tetraquarks in supersymmetric LFHQCD. [...] indicates a diquark cluster in a total quark spin $S=0$ state, (...) a cluster in a $S=1$ state. The symbol $s$ in the second column refers to the open strangeness content. All diquarks are assumed to be in the lowest possible state. }
\medskip
\renewcommand{\tabcolsep}{4pt}
{\footnotesize
\begin{tabular}{cccc|cc|cc}
\hline
\hline
q-cont. &$s$ & $I$ & $J^{PC}$& $n+L$& Mass & $n+L$& Mass  \\
&&&&&[GeV]&&[GeV] \\ \hline 
$[\overline{ud}][ud]$&0&0&$L^{(-1)^L\,(-1)^L}$&0&1.10&1&1.52\\
$[\overline{ud}](qq)$&0&1&$(L+1)^{(-1)^L\,\pm}$&0&1.33&1&1.70\\
$[\overline{qs}][qs]$&0&0,1&$L^{(-1)^L\,(-1)^L}$&0&1.42&1&1.76\\
$[\overline{sq}](sq)$&0&0,1&$(L+1)^{(-1)^L\,\pm}$&0&1.60&1&1.91\\
\hline
$[\overline{ud}][sq]$&1&$\half$&$L^{(-1)^L}$&0&1.23&1&1.61\\
$[\overline{ud}](sq)$&1&$\half$&$(L+1)^{(-1)^L}$&0&1.43&1&1.77\\
$[\overline{sq}](qq)$&1&$\half ,\textstyle{\frac{3}{2}}$&$(L+1)^{(-1)^L}$&0&1.43&1&1.77\\
$[\overline{sq}](ss)$&1&$\half$&$(L+1)^{(-1)^L}$&0&1.60&1&1.91\\
\hline
$[\overline{ud}](ss)$&2&$0$&$(L+1)^{(-1)^L}$&0&1.60&1&1.91\\
\hline
\hline
\end{tabular}}

\end{table}

If the meson in the quadruplet consists of two heavy quarks, e.g., $c$ and $\bar c$,
we have two possible tetraquarks in the quadruplet, one with 
hidden~\cite{Nielsen:2018uyn,Nielsen:2018ytt} or  one with open~\cite{Zou:2018eam} charm, see Figure \ref{fig4}, B, where the large circles represent a charm or beautiful quark. In the first case, the heavy quark of the baryon is transformed in a heavy anti-diquark, in the latter case the light quark into a light diquark. Some of the new heavy bosons can in this way be explained as tetraquarks.
In Table \ref{tab2} some candidates for hidden charm or beauty are displayed. 
\begin{table}
\caption{\lb{tab2}Some candidates for multiplets with hidden charm or hidden beauty.}
{\small \renewcommand{\arraystretch}{1.2}
\renewcommand{\tabcolsep}{2pt}
\bec 
\begin{tabular}{ccccccc}
\hline
\hline
Hadron& q-cont. &$n,L$& $J^{PC}$& $M_{theo}$& Cand. & Mass  \\
&&&&[GeV]&&[GeV] \\
 \hline
 \multicolumn{7}{c}{$c \;\bar c$}\\ \hline
Tetraq.&$[\overline{cq}](cq)$&$0,0$&$1^{++} $&3.87&$\ch_{c1}(3872)$&3.872\\
&&&$1^{+-} $&3.87&$Z_c(3900)$&3.886\\
\hline
Baryon&$(cq)c$&$0,0$&$\textstyle{\frac{3}{2}}^+$&3.76&$\Xi_{cc}\;${\footnotesize  $\half$ or $\thalf$}&3.621\\
\hline
Meson&$(\bar c\,c)$&0,1&$2^{++}$& 3.67&$\ch_{c2}(1P)$&3.556\\

 \multicolumn{7}{c}{$b \;\bar b$}\\ \hline
Tetraq.&$[\overline{bq}](bq)$&$1,0$&$1^{+} $&10.65&$Z_b(10610)$&10.610\\
\hline
Baryon&$(bq)b$&$1,0$&$\textstyle{\frac{3}{2}}^+$&10.46&$\Xi_{bb}(?)$&?\\
\hline
Meson&$(\bar b\,b)$&1,1&$2^{++}$&10.34&$\ch_{b2}(2P)$&10.268\\
\hline \hline 
\end{tabular}\enc }
\end{table}

For states with open beauty the mass of the lowest lying tetraquark could be below the threshold for a strong decay, as predicted by potential 
models~\cite{Ader:1981db,Karliner:2017qjm}.
As can be seen from Table \ref{tab3}, the existence of such states is also predicted by supersymmetric LFHQCD, namely a tetraquark with two $b$-quarks and one with a $b$ and a $c$ quark. The predicted masses are 570 and 340 MeV below the strong decay threshold, respectively. The tetraquark with open charm is predicted to be very near the threshold, so with the estimated uncertainty of $\pm 180$ MeV no conclusion can be drawn.

\begin{table}
\caption{ \lb{tab3} Masses of hadrons containing two heavy quarks. The two last columns show the lightest strong decay channel and its threshold. }
{\small
\renewcommand{\tabcolsep}{2pt}
\begin{tabular}{ccc|ccc|ccccc}
\hline
\hline
\multicolumn{3}{c|}{Mesons, $L_M=1$}&
\multicolumn{3}{c|}{Baryons, $L_B=0$}&
\multicolumn{5}{c}{Tetraquark, $L_T=0$}\\
q-cont&$J^{PC}$&$\underset{\rm [MeV]}{\rm name}$&q-cont&$J^P$&$\underset{\rm [MeV]}{\rm name}$&
q-cont&$J^P$&$\underset{\rm [MeV]}{\approx M}$ &decay&$\underset{\rm [MeV]}{\rm thresh.}$\\
\hline
$c \bar s $&$ 1^+$&$ D_{s1}(2460) $&$csq  $&$ \half^+$&$ \Xi_c(2467) $&$ cq \overline{sq}$&$  0^+$&{2550}&$D_s\pi $&$2048$ \\
$c \bar s $&$2^{+} $&$ D_{s2}^*(2569) $&$csq  $&$\thalf^+ $&$ \Xi_c(2645) $&$cq\overline{sq}$&$1^+ $&{2730}&$D_s^* \pi ..  $&$2250$ \\ \hline

$c \bar c $&$ 1^{+-}$&$ h_c(3525) $&$ccq  $&$ \half^+$&$ \Xi_{cc}(3614) $&$ cq\overline{cq} $&$  0^+$&{3660}&$\et_c\pi\pi.. $&$3270$ \\
$c \bar c $&$2^{++} $&$ \ch_{c2}(3565) $&$ccq  $&$\thalf^+ $& {$\Xi_{cc}(3770) $}&$c c \overline{qq} $&$1^+ $&{$3870$} &$D^* D.. $&{$3880^{(!)}$ }\\ \hline

$b\bar b $&$ 1^{+-}$&$ h_b(9899) $&$bbq  $&$ \half^+$&{$\Xi_{bb}(9830)$}&$ b  q \overline{b q}$&$  0^+$&{$10020 $}&$\et_b\pi\pi.. $&$9680$ \\
$b\bar b $&$2^{++} $&$ \ch_{b2}(9912) $&$bbq  $&$\thalf^+ $&{$\Xi_{bb}(10040)$}&$bb\overline{qq} $&$1^+ $&{$10230$}&$B^* B.. $&{$10800^{(!)}$} \\ \hline

$b \bar c $&$ 1^{+}$&{6550} &$bcq  $&$ \half^+$&{6660}&$ bc\overline{qq} $&$  0^+$&{6810}&$BD.. $&{$7150^{(!)}$} \\
\hline
\hline
\end{tabular}}
\end{table}

For tetraquarks containing four heavy quarks, not only conformal symmetry but also supersymmetry in the multiplets is strongly broken; the application of the formalism is therefore very questionable. Nevertheless, the mass formula \req{masst} can be applied to those states without any additional free parameter if we insert in \req{la-q} for $M_M$ the sum of the quark masses. The results are displayed in Table \ref{tab4}. It should be noted that spin and statistics demands that the lightest diquark cluster, consisting of quarks of the same flavor, must have total quark spin 1. Curiously the mass of the newly discovered $X(6900)$~\cite{Aaij:2020fnh} fits very well with the $L=0, S=2$ tetraquark state.

\begin{table}
\caption{\lb{tab4}Tentative assignment of tetraquark masses according for four heavy quarks.}
\begin{tabular}{l|c|c|c}
\hline
\hline
&$cc\bar c \bar c$ &$bb \bar b\bar b$&$cc\bar b \bar b$\\ 
$L=$0, $S=0$&6.470&19.110&12.830\\
$L=$0, $S=1$&6.680&19\,340&13.060\\
$L=$1, $S=0 $\; &6.880&19.570&13.280\\
or $ L=$0, $S=$2&&&\\
$L=1$, $S=$1&7.080&19.790&13.490\\
\hline
\hline
\end{tabular}

\end{table}

\section{Final remarks and conclusions}

There has been speculation on the existence of exotic quark 
states~\cite{Roy:2003hk} since the establishment of the quark model. Supersymmetric light-front holographic QCD predicts the existence of these states and the dependence of their masses on their quantum numbers without introducing new parameters.
The accuracy of the mass predictions is limited to $\approx 120$ MeV for states composed of light quarks, and $\approx 180$ for those containing heavy quarks.  Although the values of the masses are given by the theory, there remains, however, the problem of particle mixing, since the exotic states can mix with the conventional states without violating the Okubo-Zweig-Iizuka rule. But also in more elaborate treatments, the results of supersymmetric LFHQCD could be a valuable input.

Whereas tetraquarks are a consequence of supersymmetric LFHQCD, pentaquarks are not; they are, however, not excluded. In supersymmetric LFHQCD the pentaquark could be the member of a new supermultiplet consisting of a tetraquark, a pentaquark with two chiralities, and a hexaquark.

\section{Acknowledgements}

This work is supported in part by the Department of Energy, Contract DE--AC02--76SF00515 and by the National Natural Science Foundation of China (Grant 11805242).

\end{document}

%% file: lan-def.tex
\usepackage{graphicx}
\usepackage{epsfig,color}
\usepackage{amsmath}

\newcommand{\half}{{\textstyle{\frac{1}{2}}}}
 \newcommand{\thalf}{{\frac{3}{2}}}

\newcommand{\beq}{\begin{equation}}
\newcommand{\enq}{\end{equation}}
\newcommand{\beqa}{\begin{eqnarray}}
\newcommand{\beqast}{\begin{eqnarray*}}
\newcommand{\enqa}{\end{eqnarray}}
\newcommand{\enqast}{\end{eqnarray*}}
\newcommand{\nn}{\nonumber}
\newcommand{\req}[1]{(\ref{#1})}
\newcommand{\lb}{\label}

\newcommand{\pa}{\partial}

\newcommand{\bec}{\begin{center}}
\newcommand{\enc}{\end{center}}
\newcommand{\beqo}{\begin{quote}}
\newcommand{\enqo}{\end{quote}}

\newcommand{\cL}{{\cal L}}

\newcommand{\cR}{{\cal R}}

\newcommand{\de}{\delta}

\newcommand{\et}{\eta}

\newcommand{\la}{\lambda}

\newcommand{\ph}{\phi}

\newcommand{\ch}{\chi}
\newcommand{\ps}{\psi}

\newcommand{\De}{\Delta}

\newcommand{\Ph}{\Phi}

\newcommand{\Ps}{\Psi}